\documentclass[12pt]{article}
\begin{document}
\begin{titlepage}

\title{The $d'$ dibaryon in the quark-delocalization, color-screening
model}

\author{
 Jialun Ping\thanks{\small \em E-mail: jlping@pine.njnu.edu.cn}\\ 
{\small \em Department of Physics, Nanjing Normal University,}\\
{\small \em  Nanjing, 210097, P.R. China}\\%}
%\author{
Fan Wang\thanks{\small \em E-mail: fgwang@chenwang.nju.edu.cn}\\ 
{\small \em Center for Theoretical Physics and Department of Physics, }\\
{\small \em Nanjing University, Nanjing, 210093, P. R. China}\\%}
%\author{
T. Goldman\thanks{\small \em E-mail: tgoldman@lanl.gov}\\
{\small \em Theoretical Division, Los Alamos National Laboratory, }\\
{\small \em Los Alamos, NM 87545, USA}
}

\date{July 7, 2000}
\maketitle

\begin{abstract}
We study the questions of the existence and mass of the proposed $d'
(IJ^P=00^-)$ dibaryon in the quark-delocalization, color-screening
model (QDCSM). The transformation between physical and symmetry bases
has been extended to the cases beyond the SU(2) orbital symmetry. Using
parameters fixed by baryon properties and $NN$ scattering, we find a
mild attraction in the $IJ^P=00^-$ channel, but it is not strong enough
to form a deeply bound state as proposed for the $d'$ state. Nor does
the (isospin) I=2 N$\Delta$ configuration have a deeply bound state.
These results show that if a narrow dibaryon $d'$ state does exist, it
must have a more complicated structure.

{\noindent 12.39.-x, 14.20.Pt, 13.75.Cs}
\end{abstract}

\vspace{-8.0in}
\begin{flushright}{LA-UR-00-2797}\\
\vspace{-0.05in}
{nucl-th/0006036}\end{flushright}
\vspace{8.0in}

\end{titlepage}
 
\section{Introduction} 
Quantum chromodynamics (QCD) has been accepted as the fundamental
theory of the strong interaction. Understanding the low energy behavior
of QCD and the nature of the strong interactions of matter, however,
remains a challenge. Lattice QCD has provided numerical results
describing quark confinement between two static colorful quarks, a
preliminary picture of the QCD vacuum and the internal structure of
hadrons in addition to a phase transition of strongly interacting
matter. Phenomenological quark model analyses of hadron spectroscopy
have also provided useful physical information. However, the color
structures available in $q\bar{q}$ and $q^3$ systems are limited.
Multiquark systems involve more complicated color structures which can
not be studied directly in meson and baryon systems.  A simple example
is given by three gluon exchange\cite{lsg} and the three body instanton
interaction\cite{met} both of which do not contribute within a
colorless meson or baryon but do contribute to a multiquark system.
Therefore multiquark systems are indispensable for the full study of
the low energy behavior of QCD and the structure of strongly
interacting matter.

This report is limited to dibaryon or $q^6$ systems. Since Jaffe
predicted the H particle\cite{jaffe}, the study of dibaryons has waxed
and waned. At the end of the 1970's and beginning of the 1980's, many
dibaryon states were predicted based on the MIT bag model and some of
them were even claimed to have been observed experimentally. However
further measurement has almost completely dismissed all of them.  Most
quark models naturally find dibaryon states, except the recent model
proposed by Glozman {\em et al}~\cite{Glozman}, which has
none\cite{Stancu}.

Experimental signals have been scarce, at best.  The search for the H
particle has been continued for more than twenty years with no
indication of its existence.  Lomon predicted a high mass NN I=1
$^1S_0$ resonance at around 2.7 GeV\cite{llg} which does seem to be
supported by SATURNE pp scattering data\cite{lehar}. The
Moscow-T\"ubingen-Warsaw-Uppsala collaboration\cite{bcs} has claimed a
narrow dibaryon resonance, $d'$, centered at 2.06 GeV with a small
width of 0.5 MeV; the preferred quantum numbers are $IJ^P=00^-$.  In
contrast to all the other cases, the $d'$ dibaryon, with a mass as
small as 2.06 GeV and $IJ^P=00^-$, is hard to accommodate by the
available quark models\cite{bwf}.  Newer experiments using simple
systems have not confirmed the existence of a $d'$ signal\cite{ppexp}.

A new quark model, the quark-delocalization, color-screening model
(QDCSM), has been developed with the aim of understanding the well
known similarities between nuclear and molecular forces despite the
obvious energy and length scale differences\cite{wwdg}. The model
predicts a small mass narrow dibaryon resonance $d^*$ with $IJ^P=03^+$,
M$\sim$2.1 GeV, ${\Gamma}(NN) \sim$ 1 MeV\cite{gmww}. Although the
model has not been applied to the study of baryon resonances above the
ground state flavor octet and decuplet, where other models have shown
good accuracy\cite{CG}, we have found that it tends to underestimate
absolute orbital excitation energies. However, since our calculations
are structured to self-consistently compare dibaryon states to our
calculated two baryon thresholds, this should not lead to an
underestimate of the $d'$ dibaryon mass, within the model.

This paper reports the results of our study of the $d'$ dibaryon using
the QDCSM.  Sect.2 gives a brief description of the Hamiltonian and
wave function in QDCSM; Sect.3 describes the calculation method. For
this, a very useful extension of the transformation between the
physical bases and symmetry bases beyond the SU(2) orbital symmetry is
presented. The results are presented in Sect.4. The final section
provides a summary.

\section{Model Hamiltonian and wave function} 
The details of the QDCSM can be found in Ref.\cite{wwdg,gmww}. Here we
present only the model Hamiltonian and wave functions used in the
calculation.

The Hamiltonian for the 3-quark system is the same as the usual potential
model. For the six-quark system, it is assumed to be
\begin{eqnarray} 
H_6 & = & \sum_{i=1}^6 (m_i+\frac{p_i^2}{2m_i})-T_{CM} +\sum_{i<j=1}^{6} 
    \left( V_{ij}^c + V_{ij}^G \right) ,   \nonumber \\ 
V_{ij}^G & = & \alpha_s \frac{\vec{\lambda}_i \cdot \vec{\lambda}_j }{4} 
 \left[ \frac{1}{r_{ij}}-\frac{\pi \delta (\vec{r})}{m_i m_j} 
 \left( 1+\frac{2}{3} \vec{\sigma}_i \cdot \vec{\sigma}_j \right) \right], 
    \label{hamiltonian} \\
V_{ij}^c & = & -a_c \vec{\lambda}_i \cdot \vec{\lambda}_j 
\left\{ \begin{array}{ll} 
 r_{ij} & 
 \qquad \mbox{if }i,j\mbox{ occur in the same baryon orbit}, \\ 
 \frac{1 - e^{-\mu r_{ij}} }{\mu} & \qquad
 \mbox{if }i,j\mbox{ occur in different baryon orbits}, 
 \end{array} \right. \nonumber
\end{eqnarray} 
where all the symbols have their usual meaning except the confinement
potential $V^C_{ij}$ which will be explained below. 

The wave function of the six-quark system is written as
\begin{equation} 
|\Psi_{6q}\rangle = {\cal A} [\Psi_{B_1}\Psi_{B_2}]^{[\sigma]IJ}_{W_cM_IM_J} 
\label{wf6q} 
\end{equation} 
where $\Psi_{B_i}$ is the 3-quark system wave function,  
\begin{equation} 
\Psi_{B_1} = [ [\psi_L(1)\psi_L(2)\psi_L(3)]^{l_1} 
      \eta_{I_1S_1}]^{j_1}_{m_{j_1}}\chi_c(123) 
\label{wf3q1} 
\end{equation} 
and
\begin{equation} 
\Psi_{B_2} = [ [\psi_R(4)\psi_R(5)\psi_R(6)]^{l_2} 
      \eta_{I_2S_2}]^{j_2}_{m_{j_2}}\chi_c(456) .
\label{wf3q2} 
\end{equation} 
The single particle orbital wave functions are delocalized, as
\begin{eqnarray} 
\psi_L & = & (\phi_L + \epsilon_s \phi_R)/N_s,  \nonumber \\ 
\psi_R & = & (\phi_R + \epsilon_s \phi_L)/N_s,  \nonumber \\  
\phi_L & = & \left(\frac{1}{\pi b^2}\right)^{3/4} e^{-\frac{1}{2b^2}  
         \left(\vec{r}+\vec{s}_0/2\right)^2}, \label{wf1s} \\ 
\phi_R & = & \left(\frac{1}{\pi b^2}\right)^{3/4} e^{-\frac{1}{2b^2}  
         \left(\vec{r}-\vec{s}_0/2\right)^2}, \nonumber \\ 
N_s & = & [1+\epsilon_s^2 +2 \epsilon_s e^{s_0^2/4b^2} ]^{1/2} \nonumber  
\end{eqnarray} 
if the particle is in s-wave and as  
\begin{eqnarray} 
\psi_L & = & (\phi_L + \epsilon_p \phi_R)/N_p,  \nonumber \\ 
\psi_R & = & (\phi_R + \epsilon_p \phi_L)/N_p,  \nonumber \\  
\phi_L & = & \left(\frac{1}{\pi b^2}\right)^{3/4} \frac{\sqrt{2}}{b} 
         |\vec{r}+\vec{s}_0/2|Y_{1m} e^{-\frac{1}{2b^2}  
         \left(\vec{r}+\vec{s}_0/2\right)^2}, \label{wf1p} \\ 
\phi_R & = & \left(\frac{1}{\pi b^2}\right)^{3/4} \frac{\sqrt{2}}{b}  
         |\vec{r}-\vec{s}_0/2|Y_{1m} e^{-\frac{1}{2b^2}  
         \left(\vec{r}-\vec{s}_0/2\right)^2}, \nonumber \\ 
N_s & = & [1+\epsilon_p^2 +2 \epsilon_p (1-\frac{s_0^2}{2b^2})  
        e^{s_0^2/4b^2} ]^{1/2} \nonumber  
\end{eqnarray} 
if the particle is in p-wave, where $\vec{s}_0$ is the separation
of the two clusters. 
 
Although we use a potential model language in the description of the
confinement potential $V_{ij}^C$ in Eq.(1), our calculations employ an
extended effective matrix element method. Even though understanding of
quark confinement is limited, the use of a two body interaction to
describe the quark confinement may well be highly oversimplified,
especially for multiquark systems. For example, the three-gluon and
three-body instanton interactions mentioned above\cite{lsg,met} cannot
be expressed in terms of two-body interactions and the full
nonperturbative QCD interaction contains additional varieties, such as
condensates, which can not be expressed in terms of two body
interactions either.  Our confinement "potential" as defined in Eq.(1)
has the meaning of the usual interaction potential only in the
asymptotic region where the overlap of the $\phi_L$ and $\phi_R$
wave function orbitals is negligible. In the interaction region, it
defines a recipe for determining the effective matrix elements of the
Hamiltonian of a six quark system. (More details will be provided in
the next section.) Our goal is to include the major portion of
nonperturbative many-body quark interactions by this method.

We refer to the six-quark wave function defined in Eq.(\ref{wf6q}) as
the physical basis because it has an obvious physical meaning in terms
of a pair of baryons. In our channel coupling calculation, all of the
colorless p-wave excitation baryon-baryon channels compatible with the
$d'$ quantum numbers $IJ^P=00^-$ are included. Although the six-quark
wave function, Eq.(\ref{wf6q}), appears to have the form of $q^3-q^3$
clusters, in fact the other clusters such as $q^6, q^5q$ and $q^4q^2$
have been included because of the quark delocalization. Hidden color
channels have been omitted because we do not know anything about the
mass and interaction of colorful baryons and because a hidden color
channel can be expressed in terms of a sum of colorless
channels\cite{wwp}.

\section{Calculation method} 
 
The physical basis used in Eq.(\ref{wf6q}) is not convenient for the
calculation of matrix elements of the six-quark Hamiltonian. In order
to simplify the calculation, the physical basis (cluster basis) is
expanded in terms of symmetry bases first. Next, we use the powerful
fractional parentage (fp) expansion method to calculate the matrix
elements between symmetry basis components of six-quark system.
Finally, the matrix elements between physical basis components are
obtained by reversing the transformation between the physical and
symmetry bases.

The transformation between the physical basis and symmetry basis has
been studied by Harvey\cite{Harvey}, Chen\cite{Chen}, and
ourselves\cite{wpg} in the case where the orbital symmetry of the
three-quark cluster was limited to $[3]$ and $SU^x(2)$ orbital symmetry
was assumed. In the case of interest here, there is a p-wave quark with
respect to the right- or left-center, in addition to right- and
left-centered s-wave quarks. The orbital symmetry group needs to be
extended to $SU^x(4)$. (Only one state among the three p-wave states is
included in the calculation of the transformation coefficients, since
the spin-orbit coupling is a trivial one.) The orbital symmetry of
individual baryons should include both $[3]$ and $[21]$ configurations.
This requires and extension of the transformation method studied
previously. Here we develop two new methods to calculate the
transformation coefficients.

In the symmetry basis, the group-chain classified basis is denoted by
\begin{equation} 
|\Phi_S \rangle = \left|  \begin{array}{c} [\nu] W_x \\ ~[\sigma]  
W_c [\mu] IM_ISLJM_J   \end{array} \right\rangle ,   \label{sb} 
\end{equation} 
where $[\nu], [\sigma]$ and $[\mu]$ represent the symmetries of
orbital, color and spin-isospin degrees of freedom, and $W_x, W_c$ are Weyl
tableaux for $[\nu]$ and $[\sigma]$. Other symbols have their usual
meanings.  The group chain we use here is
\begin{equation} 
SU^{xc\tau \sigma}(36) \supset SU^x(3) \times SU^{c \tau \sigma}(12)  
 \supset SU^c(3) \times SU^{\tau \sigma}(4) \supset  
 SU^{\tau}(2) \times SU^{\sigma} (2) . 
\end{equation} 
 
The physical basis is constructed from two 3q states, 
\begin{equation} 
|\Psi_P \rangle = {\cal A}[B_1B_2]^{[\sigma]ISLJ}_{W_cM_IM_J} , 
\label{pb} 
\end{equation} 
where $B_1$ and $B_2$ represent the wave functions of two three-quark
clusters.  The physical basis defined here is a little different from
the $\Psi_{6q}$ defined in Eq.(\ref{wf6q}). The two bases are related
by Racah coefficients.  Fortunately, for the quantum numbers of $d'$,
all of the coefficients equal unity. From now on, we replace
$\Psi_{6q}$ by $\Psi_P$.

The physical basis can be expanded in terms of the symmetry basis as
\begin{equation} 
|\Psi_P \rangle = \sum_{\nu W_x \mu} C_{p-s} |\Phi_S \rangle  .  
  \label{ptos}  
\end{equation} 
By using the unitary condition of the symmetry basis, the expansion  
coefficients can be expressed as, 
\begin{equation} 
C_{p-s}= \langle \Phi_S | \Psi_P \rangle  . \label{cps} 
\end{equation} 
 
On the other hand, the six-quark symmetry basis can be expanded into two  
three-quark clusters by application of the fractional parentage technique: 
\begin{eqnarray} 
| \Phi_S \rangle & = & \sum_{1,2}  
  C^{[1^6], [\nu][\tilde{\nu}]}_{[1^3][\nu_1][\tilde{\nu}_1], 
  [1^3][\nu_2][\tilde{\nu}_2]} 
    C^{[\tilde{\nu}], [\sigma][\mu]}_{[\tilde{\nu}_1][\sigma_1][\mu_1], 
  [\tilde{\nu}_2][\sigma_2][\mu_2]} 
    C^{[\mu], IS}_{[\mu_1]I_1S_1, [\mu_2]I_2S_2} 
    C^{[\nu]W_x}_{[\nu_1]W_{x_1}, [\nu_2]W_{x_2}}  \nonumber \\   
& &  C^{[\sigma]W_c}_{[\sigma_1]W_{c_1}, [\sigma_2]W_{c_2}}  
    C^{LM_L}_{L_1M_{L_1}, L_2M_{L_2}} 
    C^{JM_J}_{SM_S, LM_L}         
    C^{IM_I}_{I_1M_{I_1}, I_2M_{I_2}} 
    C^{SM_S}_{S_1M_{S_1}, S_2M_{S_2}} \nonumber \\ 
& & \left|  \begin{array}{c} [\nu_1] W_{x_1} \\  
  ~[\sigma_1]W_{c_1}[\mu_1]I_1S_1M_{I_1}M_{S_1}L_1M_{L_1} 
  \end{array} \right\rangle  
\left|  \begin{array}{c} [\nu_2] W_{x_2} \\  
  ~[\sigma_2]W_{c_2}[\mu_2]I_2S_2M_{I_2}M_{S_2}L_2M_{L_2} 
  \end{array} \right\rangle . \label{eq2} 
\end{eqnarray} 
where $1,2$ stand for $\nu_1, \nu_2, W_{x_1}, W_{x_2}, \mu_1, \mu_2,
...$, etc. The last line is just the wave functions of two three-quark
clusters. Combining with the relevant Clebsch-Gordan (CG) coefficients,
we have
\begin{eqnarray} 
 | \Phi_S \rangle & = & \sum_{1,2}  
  C^{[1^6], [\nu][\tilde{\nu}]}_{[1^3][\nu_1][\tilde{\nu}_1], 
  [1^3][\nu_2][\tilde{\nu}_2]} 
    C^{[\tilde{\nu}], [\sigma][\mu]}_{[\tilde{\nu}_1][\sigma_1][\mu_1], 
  [\tilde{\nu}_2][\sigma_2][\mu_2]} 
    C^{[\mu], IS}_{[\mu_1]I_1S_1, [\mu_2]I_2S_2} 
    C^{[\nu]W_x}_{[\nu_1]W_{x_1}, [\nu_2]W_{x_2}} \nonumber \\ 
 & &  [B_1B_2]^{[\sigma]ISLJ}_{W_cM_IM_J}.  \label{eq3} 
\end{eqnarray}
We next apply the inter-cluster antisymmetrization operator
$$
{\cal A}= \sqrt{\frac{1}{20}} 
        \sum_{P:\mbox{inter-cluster permutations}} (-)^P P
$$
to Eq.(\ref{eq3}). Because of the total antisymmetrization of the
six-quark symmetry basis, when the operator ${\cal A}$ is applied to
the left-hand side of Eq.(\ref{eq3}), it produces 20 copies. In this
way, we obtain
\begin{eqnarray}
& & \sqrt{20} \left|  \begin{array}{c} [\nu] W_x \\ ~[\sigma ] W_c [\mu]
 IM_ISLJM_J   \end{array} \right\rangle \nonumber \\
& = & \sum_{1,2} 
  C^{[1^6], [\nu][\tilde{\nu}]}_{[1^3][\nu_1][\tilde{\nu}_1],
  [1^3][\nu_2][\tilde{\nu}_2]}
    C^{[\tilde{\nu}], [\sigma][\mu]}_{[\tilde{\nu}_1][\sigma_1][\mu_1],
  [\tilde{\nu}_2][\sigma_2][\mu_2]}
    C^{[\mu], IS}_{[\mu_1]I_1S_1, [\mu_2]I_2S_2}
    C^{[\nu]W_x}_{[\nu_1]W_{x_1}, [\nu_2]W_{x_2}} 
 {\cal A}[B_1B_2]^{[\sigma]ISLJ}_{W_cM_IM_J}.  \label{eq4}
\end{eqnarray}
Substituting Eq.(\ref{eq4}) into Eq.(\ref{cps}), and making use of the
orthonormal property of the physical bases, we obtain the expression
for the transformation coefficients,
\begin{equation}
C_{p-s} = \sqrt{20}
 C^{[1^6], [\nu][\tilde{\nu}]}_{[1^3][\nu_1][\tilde{\nu}_1],
  [1^3][\nu_2][\tilde{\nu}_2]} 
 C^{[\tilde{\nu}], [\sigma][\mu]}_{[\tilde{\nu}_1][\sigma_1][\mu_1],
  [\tilde{\nu}_2][\sigma_2][\mu_2]}
 C^{[\mu], IS}_{[\mu_1]I_1S_1, [\mu_2]I_2S_2}
 C^{[\nu]W_x}_{[\nu_1]W_{x_1}, [\nu_2]W_{x_2}} \label{eq5}
\end{equation}

This expression differs from the transformation coefficients obtained
before \cite{Chen,wpg}. It has two "new" factors:  $ C^{[1^6],
[\nu][\tilde{\nu}]}_{[1^3][\nu_1][\tilde{\nu}_1],
[1^3][\nu_2][\tilde{\nu}_2]} $ and $C^{[\nu]W_x}_{[\nu_1]W_{x_1},
[\nu_2]W_{x_2}} $. In the limit of $SU^x(2)$ with
$[\nu_1]=[\nu_2]=[3]$, the product of these two factors is a constant
with the value, $\sqrt{\frac{1}{20}}$, which returns us to the
expression of Ref.\cite{Chen,wpg}.

Recently Chen\cite{Chenpri} also proposed a method to generalize the
transformation between physical and symmetry bases. Our new method is
based on his expressions in Eqs.(9-20a) of Ref.\cite{Chenbook},
\begin{equation}
{\cal A} \left[ \left| \begin{array}{c} [\nu_1]a^{f_1} \\
 \sigma_1 \mu_1 I_1S_1 \end{array} \right)_{\omega_0^1}
 \left| \begin{array}{c} [\nu_2]b^{f_2} \\
 \sigma_2 \mu_2 I_2S_2 \end{array} \right)_{\omega_0^2} 
\right]^{[\sigma]IS} = \sum_{\tilde{\nu} \mu}
 C^{[\tilde{\nu}], [\sigma][\mu]}_{[\tilde{\nu}_1][\sigma_1][\mu_1],
  [\tilde{\nu}_2][\sigma_2][\mu_2]}
 C^{[\mu], IS}_{[\mu_1]I_1S_1, [\mu_2]I_2S_2}
  \left| \begin{array}{c} [\nu]a^{f_1}b^{f_2} \\
 \sigma \mu IS \end{array} \right),   \label{c9-20a}
\end{equation}
To extend this to the case fo interest here, first set $a^{f_1}$ to
$W_{x_1}$, $b^{f_2}$ to $W_{x_2}$. The symmetry basis used in
Eq.(\ref{c9-20a}) is not the one which we defined previously. Rather,
it is an $SU_4 \supset SU_2 \times SU_2$ irreducible basis, which is a
non-standard $SU_4$ basis, with respect to the orbital symmetry and
must be expanded in terms of the standard Gel'fand bases of $SU_4$ and
the Yamanouchi bases of $S_6$. It is only in the special $SU^x(2)$ case
that the Weyl tableau, $W_x$, is automatically fixed by the orbital
symmetry $[\nu]$. In general a further transformation from non-standard
to standard $SU_n$ bases is needed.
\begin{equation}
\left| \begin{array}{c} [\nu] \\ m \end{array}
 \begin{array}{c} [\nu_1] [\nu_2] \\ W_{x_1}W_{x2} \end{array}
\right\rangle = \sum_{W_x} 
\left| \begin{array}{c} [\nu] \\ m,W_x \end{array} \right\rangle
 \left\langle \begin{array}{c} [\nu] \\ W_{x} \end{array}
 [\nu], \begin{array}{c} [\nu_1] [\nu_2] \\ W_{x_1}W_{x2} \end{array}
\right\rangle .
\end{equation}
where the transformation coefficients, termed subduction coefficients
(SDC) of $SU_4$, are independent of $m$. From the SDC and isoscalar
factor tables of $SU_4$ and CG coefficient tables of $SU_2$, it is easy
to verify that the SDC used here is just the product of $\sqrt{20}$,
$ C^{[1^6], [\nu][\tilde{\nu}]}_{[1^3][\nu_1][\tilde{\nu}_1],
  [1^3][\nu_2][\tilde{\nu}_2]} $ and 
$C^{[\nu]W_x}_{[\nu_1]W_{x_1}, [\nu_2]W_{x_2}} $. So the two methods
give the same transformation coefficients as they should.

For the quantum numbers of $d'$, there are six physical channels,
(taking only color singlet baryons into account,) which are given in
the first column of Table I, where: $N$ denotes the state of nucleon
with quantum numbers $[\nu_i]=[3], I_i=\frac{1}{2}, S_i=\frac{1}{2}, 
l_i=0, j_i=\frac{1}{2}$; 
$\Delta$ denotes 
$[\nu_i]=[3], I_i=\frac{3}{2}, S_i=\frac{3}{2}, l_i=0, j_i=\frac{3}{2}$;
$N_1^*$ denotes
$[\nu_i]=[21], I_i=\frac{1}{2}, S_i=\frac{1}{2}, l_i=1, j_i=\frac{1}{2}$; 
$N_2^*$ denotes
$[\nu_i]=[21], I_i=\frac{1}{2}, S_i=\frac{3}{2}, l_i=1, j_i=\frac{1}{2}$; 
and $\Delta^*$ denotes
$[\nu_i]=[21], I_i=\frac{3}{2}, S_i=\frac{1}{2}, l_i=1, j_i=\frac{3}{2}$. 
The corresponding symmetry bases are given in the first row of Table
I.  There are thirteen bases. The reason we have a $6 \times 13$ table
is that the hidden color channels have been omitted.  The $d'$ is
assumed to be a linear combination of these six physical channels. The
combination coefficients are determined by diagonalizing the
Hamiltonian in the 6-dimensional space.

To calculate the matrix elements of the six-quark Hamiltonian between
symmetry bases, we use a fractional parentage expansion.  Then only the
two-body matrix elements and four-body overlaps are required. Details
can be found in Ref.\cite{wpg}. By using the transformation
coefficients derived above, the matrix elements of the six-quark
Hamiltonian between physical bases can be obtained straightforwardly.

In the calculation of the two body matrix elements of the confinement
interaction, we assume taht the normal confinement interaction should
be used for $\langle LL\arrowvert V^C\arrowvert LL\rangle$,
$\langle RR\arrowvert V^C\arrowvert RR\rangle$, 
$\langle LR\arrowvert V^C\arrowvert LR\rangle$, 
while the color-screened confining interaction should be used for
$\langle LR\arrowvert V^C\arrowvert RL\rangle$ and all others. Here
$$\langle LL\arrowvert V^C\arrowvert LL\rangle
 = \langle\phi_L(\vec{r_1})\phi_L(\vec{r_2})\arrowvert V_{1,2}^C
\arrowvert\phi_L(\vec{r_1})\phi_L(\vec{r_2})\rangle.$$
This recipe fixes the matrix elements of the confinement interaction
and distinguishes our extended effective matrix element approach from a
two-body potential-model approach. It reduces to the usual two-body
potential-model for a single hadron and therefore maintains the
successes of the constituent quark model for hadron spectroscopy. It
also reproduces qualitatively correct phase shifts in ten channels:
$NN~ST = 10, 01, 00, 11$, $N\Lambda~ ST = 0\frac{1}{2}, 1\frac{1}{2}$,
$N\Sigma~ST = 0\frac{1}{2}, 1\frac{1}{2}, 0\frac{3}{2}, 1\frac{3}{2}$,
with only one additional adjustable parameter, the color screening
constant $\mu$\cite{wwdg,wlpw}.  It is the only model which
demonstrates a similarity between nuclear and molecular forces. We take
these successes as an indication that this effective matrix element
approach includes significant components of correct physics.

For a preliminary study, we use the adiabatic approximation.  That is,
for each separation $s_0$, we determine the energy of the six-quark
system by the variational condition
\begin{equation}
 \frac{\partial E_6}{\partial \epsilon_l} =0, ~~~l=s,p
\label{vary}
\end{equation}
The effective potential between two baryons is obtained by
a subtraction~\cite{VE},
\begin{equation}
V_e(s_0) = E_6(s_0) - E_6(\infty). \label{ve}
\end{equation}
and the mass of the six-quark system is estimated by the formula,
\begin{equation}
M_6= m_1+ m_2+ V_e + E_0  \label{m6}
\end{equation}
where $m_1, m_2$ are the masses of the two baryons in isolation and
$E_0$ is the zero-point energy of the pair, $E_0=\frac{4
\hbar^2}{3ms_0^2}$, with $m$ the reduced mass of the two baryons.

\section{Results and discussion}
The model parameters, which are fixed by the baryon spectrum and $NN$
scattering, are:
$$
m_u = m_d = 313~ \mbox{MeV}, ~b = 0.603~ \mbox{fm}, ~a = 25.13~ 
\mbox{MeV/fm$^2$}, ~\alpha_s = 1.54, ~\mu = 1.6~fm^{-2}.
$$
Both single channel and channel coupling calculations were carried out.
The results are shown in Table II. From these results, we observe a
mildly attractive interaction between $N$ and $N^*$, $\sim 100$ MeV.
Since the delocalization is approximately 1, this is a six-quark state,
not a two-baryon state.  However this attraction is not enough to form
a deeply bound state such as $d'$. Channel coupling adds a little more
attraction ($\sim 10$ MeV), does not change this conclusion.

These results are expected in the QDCSM. From our previous study of the
effective potential between baryons~\cite{VE}, we found that as a
general trend there are strong attractions in decuplet-decuplet channels, but
the attraction in octet-octet channels is weak. The excited $N^*$ state
of the nucleon is still in a flavor octet, so the attraction between
$N$ and $N^*$ would not be expected to be strong, although the
existence of the $p$-wave quark adds a little more attraction. The mass
of $\Delta\Delta^*$ channel is much larger than that of the $NN^*$
channels, so the effect of channel-coupling to it is small.

The effect of the harmonic oscillator parameter, $b$, of the quark
orbital wave function has also been studied.  G. Wagner {\em et
al.}~\cite{Fas2} concluded that it is impossible to describe the
dibaryon, $d'$, with the same parameters as those used for single
baryons. In particular, they found it impossible to get a $d'$ mass as
low as 2.06 GeV if the same $b$ is used for both 3-quark and six-quark
systems.  To check their statement, the results obtained with a larger
value of $b$ in the six-quark calculation are also shown in Table II.
By varying $b$, the attraction between $N$ and $N^*$ could be
increased. However, if a different $b$ is used in the 3-quark and
six-quark calculations, the subtraction procedure, shown in
Eq.(\ref{ve}) to obtain the effective potential, is no longer reliable.
A dynamic calculation is needed. It is possible that a deeply bound
$d'$ state might be obtained in the QDCSM if a larger parameter $b$ is
used for the six quark system. However, the physical meaning of such a
model calculation is obscure.

The possibility that the $d'$ has isospin $I=2$ has been discussed in the
literature. We have carried out both adiabatic and dynamical
$IJ^P=20^-$ N$\Delta$ single channel calculations. With the same model
parameters, the adiabatic calculation produces a mass of 2173 MeV for
the $d'$. A dynamical calculation with eight Gaussians (spanning
cluster separation coordinate values from 0.6 to 4.8 fm) produces a mass
of 2191 MeV and one with fifteen Gaussian (spanning cluster separation
coordinate values from 0.6 to 9.0 fm) produces a mass of 2178 MeV. These
trends strongly suggest that this is a scattering state, not a bound
state. So we conclude that no state with a mass as low as 2.06 GeV can
be obtained in the $N\Delta$ channel in the QDCSM.

\section{Summary}
We have studied the proposed $d'$ state in the QDCSM using the physical
bases $NN^*$ and $\Delta\Delta^*$. Our results show that a mild
attraction does develop in the $IJ^P=00^-$ state, but this attraction
is not strong enough to form a deeply bound state with a mass as low as
2.06 GeV. All of the p-wave excitation baryon-baryon channels with
$d'$-compatible quantum numbers have been included in our channel
coupling calculations. Because the QDCSM  includes quark
delocalization, all of the configurations. $q^6, q^5q, q^4q^2$ and
$q^3q^3$, are included in our model space.

Nor do we find such a low mass state in the $I=2$ $N$$\Delta$ channel
in our model approach.

Including the $d'$ results of the Faessler group\cite{bwf,Fas2}, we
conclude that, if the $d'$ is experimentally verified as a real
dibaryon resonance, it must have a more complicated structure than
studied here, such as including $q^7\bar{q}$ components.

\vspace{0.2in}
Special thanks are due to Jin-Quan Chen for constructive discussions.
This research is supported by the National Science Foundation of China,
Fok Yingdung Educational fund, National Science Foundation of Jiangsu
Province and in part by the U.S. Department of Energy under contract
W-7405-ENG-36.

\begin{table}
\caption{The transformation coefficients between physical and symmetry
bases.}
\begin{tabular}{c|ccccccc} \hline
   & $[51]_1[321]$ & $[51]_2[321]$ & $[42]_1[51]$ & $[42]_2[51]$ & 
     $[42]_1[411]$ & $[42]_2[411]$ & $[42]_1[321]$ \\ \hline
$NN_1^*$ & $-\sqrt{\frac{4}{45}}$ & $\sqrt{\frac{1}{45}}$ & 
  $-\sqrt{\frac{4}{81}}$ & $\sqrt{\frac{16}{405}}$ & $-\sqrt{\frac{2}{81}}$ &
  $\sqrt{\frac{8}{405}}$ & $\sqrt{\frac{5}{81}}$ \\
$NN_2^*$ & $-\sqrt{\frac{1}{45}}$ & $\sqrt{\frac{1}{180}}$ & 
  $-\sqrt{\frac{4}{81}}$ & $-\sqrt{\frac{16}{405}}$ & $-\sqrt{\frac{8}{81}}$ &
  $\sqrt{\frac{32}{405}}$ & $\sqrt{\frac{5}{324}}$ \\
$\Delta\Delta^*$ & $\sqrt{\frac{4}{45}}$ & $-\sqrt{\frac{1}{45}}$ & 
  $-\sqrt{\frac{1}{81}}$ & $\sqrt{\frac{4}{405}}$ & $-\sqrt{\frac{8}{81}}$ &
  $\sqrt{\frac{32}{405}}$ & $-\sqrt{\frac{5}{81}}$ \\
$N_1^*N$ & $-\sqrt{\frac{4}{45}}$ & $-\sqrt{\frac{1}{45}}$ & 
  $\sqrt{\frac{4}{81}}$ & $\sqrt{\frac{16}{405}}$ & $\sqrt{\frac{2}{81}}$ &
  $\sqrt{\frac{8}{405}}$ & $-\sqrt{\frac{5}{81}}$ \\
$N_2^*N$ & $\sqrt{\frac{1}{45}}$ & $\sqrt{\frac{1}{180}}$ & 
  $\sqrt{\frac{4}{81}}$ & $\sqrt{\frac{16}{405}}$ & $-\sqrt{\frac{8}{81}}$ &
  $-\sqrt{\frac{32}{405}}$ & $\sqrt{\frac{5}{324}}$ \\
$\Delta^*\Delta$ & $-\sqrt{\frac{4}{45}}$ & $-\sqrt{\frac{1}{45}}$ & 
  $-\sqrt{\frac{1}{81}}$ & $-\sqrt{\frac{4}{405}}$ & $-\sqrt{\frac{8}{81}}$ &
  $-\sqrt{\frac{32}{405}}$ & $-\sqrt{\frac{5}{81}}$ \\ \hline \hline
   & $[42]_2[321]$ & $[411][411]$ & $[411][321]$ & $[321][51]$ & 
     $[321][411]$ & $[321][321]$ & \\  \hline
$NN_1^*$ & $-\sqrt{\frac{4}{81}}$ & $\sqrt{\frac{1}{27}}$ & 
  $-\sqrt{\frac{4}{54}}$ & $-\sqrt{\frac{16}{45}}$ & $\sqrt{\frac{4}{135}}$ &
  $-\sqrt{\frac{4}{27}}$ &  \\
$NN_2^*$ & $-\sqrt{\frac{1}{81}}$ & $\sqrt{\frac{4}{27}}$ & 
  $-\sqrt{\frac{1}{54}}$ & $\sqrt{\frac{16}{45}}$ & $\sqrt{\frac{16}{135}}$ &
  $-\sqrt{\frac{1}{27}}$ &  \\
$\Delta\Delta^*$ & $\sqrt{\frac{4}{81}}$ & $\sqrt{\frac{4}{27}}$ & 
  $\sqrt{\frac{4}{54}}$ & $-\sqrt{\frac{4}{45}}$ & $\sqrt{\frac{16}{135}}$ &
  $\sqrt{\frac{4}{27}}$ &  \\
$N_1^*N$ & $-\sqrt{\frac{4}{81}}$ & $-\sqrt{\frac{1}{27}}$ & 
  $\sqrt{\frac{4}{54}}$ & $-\sqrt{\frac{16}{45}}$ & $\sqrt{\frac{4}{135}}$ &
  $-\sqrt{\frac{4}{27}}$ &  \\
$N_2^*N$ & $\sqrt{\frac{1}{81}}$ & $\sqrt{\frac{4}{27}}$ & 
  $-\sqrt{\frac{1}{54}}$ & $-\sqrt{\frac{16}{45}}$ & $-\sqrt{\frac{16}{135}}$ &
  $\sqrt{\frac{1}{27}}$ &  \\
$\Delta^*\Delta$ & $-\sqrt{\frac{4}{81}}$ & $\sqrt{\frac{4}{27}}$ & 
  $\sqrt{\frac{4}{54}}$ & $\sqrt{\frac{4}{45}}$ & $-\sqrt{\frac{16}{135}}$ &
  $-\sqrt{\frac{4}{27}}$ &  \\ \hline
\end{tabular}

\vspace{0.5in}

\caption{Effective potentials and masses of six-quark systems.}
\begin{tabular}{c|cccccccc} \hline
   & $b$(fm) & $s_0$(fm) & $\epsilon_s$ & $\epsilon_p$ & 
     $E_6(s_0)$(MeV) & $E_6(\infty)$(MeV) & $V_e$(MeV) & $M_6$(MeV) \\ \hline
s.c. & 0.603 & 1.0 & 1.0 & 1.0 & 2376 & 2473 & -97  & 2466 \\
c.c. & 0.603 & 1.0 & 0.6 & 1.0 & 2364 & 2473 & -109 & 2454 \\
s.c. & 0.85 & 0.9 & 1.0 & 1.0 & 2123 &      &       &  \\
c.c. & 0.85 & 0.9 & 1.0 & 1.0 & 2121 &      &       &  \\ \hline
\end{tabular} 
\end{table}

\end{document}